\documentclass[a4paper, 11pt]{amsart}
\usepackage[a4paper, top=30mm, bottom=30mm, left=30mm, right=30mm]{geometry}
\usepackage{setspace}
\pdfoutput=1

\usepackage[T1]{fontenc}
\usepackage{lmodern}
\usepackage[american]{babel}
\usepackage[hyphens]{url}

\usepackage{datetime}
\usepackage{calc}
\usepackage{xcolor}
\usepackage{float}
\usepackage{graphicx}
\DeclareGraphicsExtensions{.pdf,.png,.jpg,.jpeg}
\usepackage{booktabs}
\usepackage{amsmath}
\usepackage{mathtools}
\usepackage{amsthm}
\usepackage{amssymb}
\usepackage{bm}
\usepackage{amsfonts}
\usepackage{mathrsfs}
\usepackage{siunitx}
\usepackage{natbib}

\usepackage{pdfpages}

\makeatletter
\def\input@path{{aux-lean-codes/}}
\makeatother

\usepackage{hyperref}
\hypersetup{
pdftitle={Statistical Theory in the Age of Machine-Assisted Mathematics},
pdfauthor={Pietro Coretto},
pdfsubject={},
pdfcreator={},
pdfproducer={},
pdfkeywords={},
breaklinks  = true,
pdfborder={0 0 0},
colorlinks=true,    
linkcolor=magenta,  
citecolor=magenta,  
filecolor=blue,     
urlcolor=blue       
}

\usepackage{enumitem}
\setlist{itemsep=2pt, parsep=2pt,  topsep=3pt, partopsep=0pt}

\theoremstyle{plain}

\theoremstyle{definition}

\theoremstyle{remark}

\newcommand{\R}{\mathbb{R}}

\newcommand{\E}{\mathbb{E}}
\newcommand{\Prob}{\mathbb{P}}
\newcommand{\Var}{\mathbb{V}ar}

\DeclareMathOperator*{\argmin}{arg\,min}
\newcommand{\convd}{\xrightarrow{\;d\;}}

\newcommand{\norm}[1]{\lVert #1 \rVert}
\newcommand{\abs}[1]{\lvert #1 \rvert}

\newcommand{\hyp}[1]{\textsf{#1}}

\newcommand{\Lean}{\textsf{Lean}}
\newcommand{\lk}[1]{\textsf{#1}}
\newcommand{\leanfont}{\ttfamily}

\let\pcorigtexttt\texttt
\renewcommand{\texttt}[1]{{\def\_{\textunderscore\allowbreak}%
  \pcorigtexttt{#1}}}

\newcommand{\leanfigure}[1]{%
  \begin{center}%
    \setlength{\fboxrule}{0.3pt}%
    \setlength{\fboxsep}{4pt}%
    \fcolorbox{black!30}{white}{\includegraphics{#1}}%
  \end{center}}

\begin{document}

\title[Statistical Theory in the Age of Machine-Assisted Mathematics]{
Statistical Theory in the Age  of \\ Machine-Assisted Mathematics
}
\author{%
Pietro Coretto%
}\thanks{%
Department of Economics and Statistics, %
University of Salerno (Italy), %
E-mail:~\texttt{pcoretto@unisa.it}} %



\maketitle

\begin{abstract}
The computational revolution is advancing at an unprecedented pace.
The combination of proof-assistant technologies and generative AI tools has recently enabled the solution of complex problems in pure mathematics at a scale that seemed unattainable only a few years ago, but they have not yet become standard tools in the development of statistical theory.
In this paper we do not present new theoretical results; instead, we discuss five case studies involving classical problems in statistics and describe how a machine proof checker changes the thought process behind their proofs, in order to stimulate reflection on how these technologies may transform theoretical research and advanced statistical education.
Statistical theory compresses mathematical content into expressions such as ``under the usual regularity conditions''; once AI writes the proofs and machines check them, the human reader is left with the statement and its hypotheses, which is exactly what such expressions compress.
Against the fear that automation erodes skills, our experience was the opposite: what changes is the job of the theorist, which moves from the steps of a proof to its statement.
The new paradigm can have positive effect also on methodologists and applied statisticians, because these tools allow for quick and precise check of formal guarantees necessary to justify the application of a specific model and/or methodology.
We draw consequences for refereeing, doctoral teaching, and we encourage a collective effort toward shared formal libraries for statistics. 
Just as high-level programming languages revolutionized empirical research, machine-assisted formalization may introduce a new paradigm for the development, verification, and communication of statistical theory.
\end{abstract}


\section{Introduction}
\label{sec:intro}
Proof assistants combined with powerful generative artificial intelligence (AI) tools have become an active part of mathematical research.
A proof assistant is a program that checks every step of a proof against a fixed set of rules; \Lean, the one most widely adopted by mathematicians, comes with Mathlib, a community library that now holds a large and growing body of machine-checked mathematics \citep{mathlib2020lean}.
Major projects have shown that formalization can engage current research and not only settled results: Scholze's liquid tensor experiment verified a foundational result in condensed mathematics \citep{scholze2022liquid}, and the Gowers--Green--Manners--Tao proof of Marton's conjecture was formalized soon after its announcement \citep{gowers2023marton}.
Related efforts have begun to reach probability, statistical learning, and asymptotic statistics \citep{zhang2026ai4slt,sonoda2025rademacher,degenne2025markov,degenne2025brownian,wei2026hypothesis}.
Theoretical statistics should therefore ask not only whether these tools are useful, but how they can change and improve our field.
 
In 2026 the question has become pressing.
In September, OpenAI announced that an internal system had proved finite-time blow-up for the Navier--Stokes equations under a smooth external force and released a \Lean\ formalization with the proof \citep{openai2026navierstokes}; in October the same company announced a proof of the quasi-Riemann hypothesis, in a first version verified in \Lean\ and in a shorter second version without formal verification \citep{kakaes2026quasirh}.
Both releases soon met the same objection: \citet[Section~3]{bastounis2026lost} argued that the \Lean\ theorem of the Navier--Stokes release establishes less than the paper claims, and the number theorists examining the quasi-Riemann proof set the verified version aside to study the shorter one \citep{kakaes2026quasirh}.
In both cases the kernel had done its part, and what remained open was what the formal statement says and who reads it; with AI, as \citet{detoffoli2026aftermath} put it, the logical notion of proof, which a kernel checks, and the intelligible one, which a mathematician follows, can come apart.
For statistics, whose open problems so often turn on long chains of technical conditions, the question is: when AI writes the proofs and machines check them, who reads the hypotheses?
This paper takes the question up from the statistician's side, as a discussion built on five classical results which we formalized, and not as a collection of new theorems.
 
The value of a proof assistant is not just certification.
Nobody needs one to believe that the normal maximum likelihood estimator is consistent.
The more interesting use is that it enforces explicit verification standards.
Statistical proofs often compress packages of topology, analysis, probability, and algebra into phrases like ``under regularity conditions'', ``by a law of large numbers'', or ``with high probability''.
This compression is efficient, but it conceals which condition supports which step: the phrase works as a citation, which names the package without recording what each step uses, and its cost is not error but non-recoverability (Section~\ref{sec:synthesis}).
A proof assistant cannot rely on that shared background, so that each condition has to enter as a named hypothesis of the step which uses it, and the result is a representation in which a claim and the assumptions that support it can be inspected together.
 
We use that property to study five results of classical statistical theory, from the consistency of the maximum likelihood estimator to the LASSO oracle inequality, formalized in \Lean~4 with Mathlib.
The formalizations are retrospective, and the point of the five cases is not the theorems, which were never in doubt, but the thought process behind their proofs and how it changes when a machine reads along: which choices the informal argument leaves open, at which step each condition is used, and what is left for the author to decide once the kernel has accepted the proof.
A version-locked copy of the code, with structured comments that trace each condition to its source, is provided in the electronic supplement for reproduction and auditing \citep{coretto2026supplement}.
 
Out of the exercise comes, first of all, a general argument: the compressed idiom is a bibliographic device rather than a proof step, and a kernel which certifies the steps leaves to the human reader exactly the conditions that the idiom compresses (Section~\ref{sec:synthesis}).
The cases supply the evidence, since in each of them the formalization ties every condition it uses to the step that applies it.
Of the 32 conditions with mathematical content, the cited sources state 22 and presuppose 9; the remaining one is the sharpest finding, a law of large numbers invoked at a random intermediate point in the score-equation proof of asymptotic normality, which the pointwise law, binding its parameter outside the almost-sure quantifier, does not reach (Section~\ref{sec:case3}).
The repair is classical \citep{ferguson1996course,newey1994large}, and empirical-process treatments bypass the expansion altogether \citep{pollard1985clt,vandervaart1996weak}; what the case yields is not a new theorem but a standard enforced on every argument.
Beyond our cases, three independent formalization projects report formal statements that departed from the intended claims, a departure which the kernel does not check \citep{wei2026hypothesis,zhang2026ai4slt,tan2026causalforge}.
Consequences follow for working methods, refereeing, doctoral teaching, and the infrastructure that a statistics-oriented library needs first (Section~\ref{sec:synthesis}).
 
Two of those consequences run against the current mood and deserve a word here.
It is feared that automated proving will erode the competence of the theorist; our experience in producing the five developments was the opposite.
Most of the effort went into fixing the framework in which each argument takes place, the parameter set, the norm, the order of the quantifiers, the measurability of an estimator, because a formal environment does not let the framework stay unspecified, and we learned more about results that we had known for years than any derivation by hand had taught us.
What changes is not the understanding that the work requires but the job itself, which moves from writing the steps of a proof to stating the claim and reading its hypotheses.
The gain, moreover, is not confined to theorists.
A library of verified regularity conditions for the standard models would let a methodologist or an applied statistician check quickly whether the model at hand falls within the scope of a theory and of the guarantees it offers, and whether a method is appropriate in a given context, a check that today rests on a phrase (Section~\ref{sec:case2}).
 
The five cases demonstrate the diagnosis within their own scope.
They were chosen to exhibit mechanisms rather than to form a representative sample, and the tiers were assigned within this project; the counts therefore describe this development and not the statistical literature, whose compression remains to be measured by audits on a corpus of textbooks and papers (Section~\ref{sec:synthesis}), and the study does not measure time either, beyond the unmeasured estimate reported there.
The broader claim remains a conjecture: once suitable libraries and practices exist, formal tools may help statisticians develop theory more reliably, reuse arguments at a finer grain, and teach the architecture of proofs more clearly, much as high-level languages such as \textsf{R} and \textsf{Python} changed empirical and computational work.
 
Section~\ref{sec:background} introduces proof assistants and Mathlib for a statistical audience and closes with a primer on reading \Lean\ code; Section~\ref{sec:cases} presents the five cases, Section~\ref{sec:synthesis} draws the lessons for research and teaching from their gap inventory and from the independent evidence, and Section~\ref{sec:conclusion} concludes.
The supplement also gives the auditing protocol and a complete user guide, and a working copy of the code is kept in a public repository.

\section{Proof Assistants for the Theoretical Statistician}
\label{sec:background}

\subsection{What is a proof checker}
\label{sec:checker}

A proof assistant is a programming language in which mathematical statements and proofs are written in the form of code and then checked by a machine.
A statement is given in a fixed formal syntax, a proof is a script which puts together a full logical derivation, and a small, independently auditable program known as the \emph{kernel} verifies the result.
The compilation will not take place if any step lacks justification, if a hypothesis is not discharged, or if a lemma is applied in a case where it does not fit, just as a program fails to compile when there is a data-type error.
The only exception is the case of an explicitly ``\emph{admitted step}'' (Section~\ref{sec:primer}); in this instance the file compiles with a warning and the kernel logs the \emph{admission}.
A thing cannot be invoked unless it has first been stated, and so a lemma must be referred to by name with all of its hypotheses discharged, and the usual expression ``by the usual argument'' has no equivalent.
Definitions also involve commitments: in order for a statement to be formulated, every notion it includes must first have a precise formal meaning assigned to it, which means that any conflation which has gone unnoticed in the prose will appear as soon as the two notions are written down.

What the kernel certifies is a conditional.
A file that compiles demonstrates that each theorem whose proof is complete reaches its conclusion from the hypotheses stated in it, via the library and the standard axioms.
The theorem's formal statement is exactly the same as the conclusion which the argument authorizes, that is to say, the conclusion obtained from the given hypotheses in the proof that has been checked.
The result is deterministic and will be the same on every machine that is running the same software.
Two judgments are not included in the certificate.
The kernel does not verify that the formal statement accurately expresses the informal claim it is intended to make, and it also does not check that the statement contains anything that is worth proving.
It also does not take into account the hypotheses; a theorem which assumes more than its source states is treated just like any other one, since the kernel checks the derivation and not the premises.
The judgments are still human, and that is the key point in the study of the five cases in Section~\ref{sec:cases}.

Deliberately, the kernel is kept small: it implements the rules of a dependent type theory and has only limited knowledge of mathematics.
The real numbers, the measures, the topology, and all the theorems relating to them are part of a library, since they are the definitions and theorems which earlier users wrote and which the kernel has already verified.
A library is a collection of checked statements external to the kernel.
A library  does not require any trust by itself, since every item in it includes a verified proof, and a development can be based on hundreds of thousands of such proofs without causing the program used to check each new proof to become any larger.
The largest library that currently exists is Mathlib.
Mathlib is based on only three axioms.
Proofs are composed using tactics, which are commands that carry out one step of the argument (for example, carrying out an induction, rewriting using this identity, or applying that lemma) from a library of patterns that have been accumulated by the community.
The kernel verifies the derivation that the tactics generate, not the tactics themselves.
The simplest exercise shows the advantage of such a level of detail: when $0+n=n$ is formalized over the natural numbers it becomes clear that this is not the same as $n+0=n$, because the latter is true by definition whereas the former has to be proved using induction \citep{tao2016analysis,tao2023analysislean}.

Consider the dominated convergence theorem.
A statistical proof refers to it in a phrase and then continues.
In \Lean\ the theorem is a library declaration, namely \texttt{MeasureTheory.tendsto\_integral\_of\_dominated\_convergence} \citep{mathlib2026dct}, and to use it one has to apply this declaration to the sequence in question.
The application passes type-checking only if the user gives four arguments: a proof that each $f_n$ is measurable, a dominating function $g$ together with a proof that $g$ is integrable, a proof that $\abs{f_n} \le g$ almost everywhere for each $n$, and a proof that $f_n$ converges to $f$ almost everywhere.
It is not until then that the kernel grants the conclusion, the convergence of $\int f_n$ to $\int f$.
It grants it for the integral as Mathlib defines it, which means that any statement concerning another integral must first be linked to that one.
A hypothesis which the user does not provide is either proved in advance or included in the hypotheses of the theorem being constructed, with all readers being able to see it.
The condition that the prose reduces to ``by dominated convergence'' breaks down into four named arguments, and the compiler indicates which one is missing.

\subsection{\Lean~4 and Mathlib}
\label{sec:leanmathlib}

The paper makes use of \Lean~4, which is a proof assistant and a general-purpose programming language based on dependent type theory \citep{demoura2021lean4}.
\Lean\ provides the formal language, the kernel, and the editor tooling.
The mathematics is provided by Mathlib \citep{mathlib2020lean}, a standalone library which has been developed in \Lean\ and is looked after by its community.
There are other mature alternatives available (such as Rocq, Isabelle/HOL, HOL Light, and others; \citealp{bertot2004coq,nipkow2002isabelle,harrison2009hollight}), and the decision between them is a pragmatic one, not an assertion of superiority.
Two facts determined our decision.
The mathematics upon which statistical theory is based--namely topology, measure theory and probability, real analysis, and algebra--is all contained within Mathlib using a single vocabulary and a common set of conventions, and since every item in it has gone through the kernel, any result that has been formalized once remains available for use in all subsequent arguments.
The various systems currently designed to produce formal mathematics--such as those which offer tactic suggestions within the editor \citep{song2024leancopilot} and those which carry out full proof search \citep{deepmind2024alphaproof}--all focus on \Lean\ rather than on any other assistant, and as a result the AI tools available at the present time offer the best support for \Lean.

An artificial-intelligence system is not a proof assistant, even though the two are now often associated with each other.
A proof checker combined with generative artificial intelligence presents a new opportunity for the development of theories based on mathematics and formalization.
What the kernel cannot do, generative models are able to do, namely translating informal mathematics into formal syntax \citep{wu2022autoformalization}, proposing tactics, and searching for complete derivations.
They alter the way in which researchers produce proofs since the model puts forward candidate derivations in amounts that no one could write out by hand, the kernel rejects each candidate that does not work, and the error message from each failure is used to guide the next attempt.
When a strategy fails, only a few minutes are lost instead of days, and a researcher is able to examine one claim using a number of different strategies at the same time.
The guarantee given in Section~\ref{sec:checker} is independent of the method used to obtain a proof, since the kernel verifies the entire derivation against the formal statement before accepting it.
The system examines each proof in relation to the statement it is given, and it is up to the reader to decide whether the statement reflects what the author intended.

Most of statistics is not yet included in Mathlib.
The only statistical component present in the snapshot used here, which was fixed to a commit from February 2026, is a decision-theoretic module that defines Bayes and minimax risk \citep{degenne2026decision}, and the statistical libraries of 2026, as reviewed below, developed outside Mathlib.
The library was put to use in three different ways in the five cases.
In the cases where Mathlib contained the result, the development made use of it, for example with the dominated parametric integral in Section~\ref{sec:case2} and Slutsky's theorem in Section~\ref{sec:case3}.
In cases where a result was absent from the library, the development was either able to prove it—taking the example of the argmax engine in Section~\ref{sec:case1}—or treated it as a named and explicit hypothesis, as in the case of the central limit theorem for the score in Section~\ref{sec:case3}.
In cases where the result was in a form which could not be used by a statistical statement, the development had to repackage it, as Section~\ref{sec:case3} shows.
The supplementary materials include the file \path|coverage_audit.lean| which records the searches, and each statement in this paper concerning what Mathlib holds or lacks refers to the pinned snapshot.
Since both Mathlib and \Lean\ are developing rapidly, any claim about what the library lacks has a short lifespan.

Work that is formal in nature and related to statistics has only recently been carried out, and a great deal of it appeared after our snapshot was taken.
In Rocq a probability library has obtained concentration inequalities \citep{affeldt2025concentration}.
In the field of statistical learning theory, \citet{zhang2026ai4slt} set out the empirical processes, the Gaussian Lipschitz concentration, and Dudley's entropy integral, while \citet{sonoda2025rademacher} established generalization bounds by means of Rademacher complexity.
In 2026 three general statistical libraries appeared.
Statlib, a project which deals with the establishment of definitions and statements prior to proofs, includes coverage of inference models, quadratic-mean differentiability, contiguity, e-values, statistical divergences, and lower bounds for hypothesis testing \citep{statlib2026}.
StatLean includes limit theorems as well as conformal prediction and states forty-four axioms, most of which are standard results that are given without proof \citep{statlean2026}.
FormalProbability includes treatments of the central limit theorem, causal identification, and estimators that have been corrected for machine-generated labels \citep{linegar2026formalprobability}.
The most relevant work in this area is that of \citet{wei2026hypothesis}, who develop asymptotic statistics for both parametric and semiparametric models using a pipeline consisting of language-model agents, and \citet{tan2026causalforge}, who combine Causalean, a causal inference library, with an agent that suggests and proves new results.
Since a compiled proof does not demonstrate that its statement expresses the claim intended, both projects include a check that the kernel cannot itself make.
Wei et al.\ check each hypothesis of the main theorem against the source text, while Tan and Syrgkanis examine each formal statement in relation to the informal claim it encodes.
The libraries are developed outside Mathlib and on six different \Lean\ toolchains, and only Causalean builds on another of them.
They direct work towards the creation of libraries and pipelines.
This work poses an alternative question: it is concerned with what can be discovered about the statistical arguments through formalization, specifically regarding the conditions that a classical proof employs, where those conditions are stated in the sources, and where the formal statement differs from the informal one.
Moreover, in this work we argues, using well known theoretical problems, that the role of a modern statistician between proof checkers and artificial intelligence technologies, is still central but probably different compared to the past. 

\subsection{Reading \Lean\ code}
\label{sec:primer}

In this section we provide some guidelines concerning the main constructs of \Lean\ in order to aid the reader when reading the code of the five case studies.
It is not meant to serve as a comprehensive guide to \Lean, but merely as an example of the key ideas.
The theorem is given together with its conclusion, without its proof, and a definition is provided along with the body which determines its meaning.

\begin{figure}[H]
\leanfigure{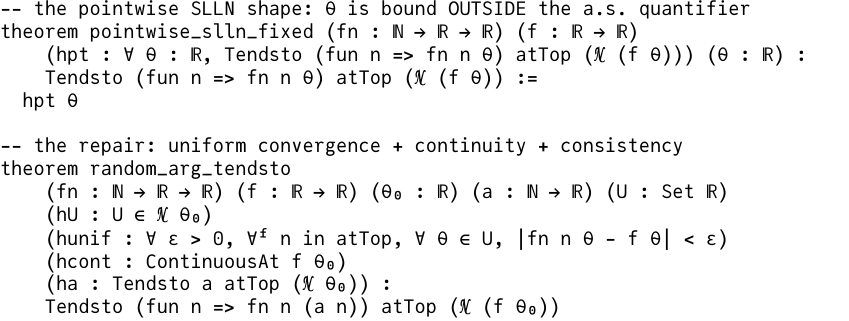}
\caption{Case~3: the pointwise law and its random-argument repair.}
\label{fig:lean-02}
\end{figure}

Consider the display in Figure~\ref{fig:lean-02}.
This fragment shows a proof, the function application that closes \texttt{pointwise\_slln\_fixed}, which is discussed in Section~\ref{sec:case3}.
A \lk{def} (definition) introduces an object and establishes its meaning; in order to use any property other than that given in the definition it must first be proved.
A \lk{theorem} consists of a statement together with its proof, and \lk{lemma} is a synonymous term for this kind of construction.
The arguments of a theorem, as listed in the parentheses preceding the final colon, bind the objects and the hypotheses it assumes.
A hypothesis is thus an argument that has a name: for example, the label \texttt{hsep} followed by the statement to which it refers.
The conclusion is the statement that comes after the final colon.
For instance, in the fragment of Section~\ref{sec:case1} the hypotheses \texttt{hsep}, \texttt{hunif}, and \texttt{hopt} state separation, uniform convergence, and the maximizing property respectively, and the conclusion is that the estimator converges.

An important keyword of \Lean\ suspends the discipline of Section~\ref{sec:checker}, namely \lk{sorry}.
The \lk{sorry} command causes any open proof obligation to be closed in the absence of an argument.
The compilation proceeds with a warning and the kernel notes the dependency: all theorems whose proofs include a \lk{sorry} depend on the built-in constant \texttt{sorryAx} as reported by \texttt{\#print axioms}.
In this paper a step which is closed either by means of \lk{sorry} or by means of its earlier synonym \lk{admit} is referred to as an \emph{admitted step}.
An \lk{axiom} declaration establishes a global assumption: it has no proof, but at each point where it is used it is as if it were a proved theorem.
The actual difference between a theorem, a lemma and an axiom is that an axiom need not have a proof to be used.

There are three points about notation that are important in what follows.
First of all, the binders are read from left to right before the final colon, which means that the order of the quantifiers is the same as the order of the binders.
Second, the limits: {\leanfont Tendsto \ensuremath{\theta}hat atTop (\ensuremath{\mathcal{N}} \ensuremath{\theta_0})} asserts that the sequence {\leanfont \ensuremath{\theta}hat} converges to {\leanfont \ensuremath{\theta_0}}, the term {\leanfont atTop} meaning ``as the index grows without bound'' and {\leanfont \ensuremath{\mathcal{N}} \ensuremath{\theta_0}} denoting the neighborhoods of {\leanfont \ensuremath{\theta_0}}, both interpreted as filters, which are a feature of Mathlib for expressing limits (for more details see Sections~\ref{sec:case1} and~\ref{sec:case3}).
Third, comments: anything that comes after \texttt{'--'} is considered a comment, and the bracketed indicators found within comments, especially \texttt{[GAP]}, are the structured audit annotations which have been specifically designed for this paper and are described in Section~\ref{sec:cases}.

\section{Five Case Studies}
\label{sec:cases}
In all five cases we compare the informal argument of the cited source with its formal version and identify each condition that is required by the formal proof together with the step at which it is introduced.
All the statements in this section concerning what a source says or what it omits refer to the source cited at the relevant point.
The supplement's Reading Guide (Section~S5) goes through the source file for each case, and Table~S1 assigns the case to the file together with its main declarations.
In Table~\ref{tab:cert-hierarchy} the results checked by the kernel are separated, case by case, from the inputs taken as explicit hypotheses and from those correspondences which exist only in the prose and not in any formal statement.
The five files contain no \lk{axiom}, \lk{sorry} or \lk{admit} declaration.
Each theorem makes use of nothing but \Lean's standard axioms, and \texttt{sorryAx} is not included in any axiom profile.

The audit lists, as a \texttt{[GAP]} comment in the source files, every condition which the formalization identified, each accompanied by an identifier, for example \hyp{H1.4}, and a tier which categorizes the condition according to its connection with the source cited for it.
The four tiers are as described by \citet{wei2026hypothesis}: \hyp{HC} (high confidence) in the case where the source states the condition either explicitly or in effect, \hyp{IMP} (implied) when the source's notation or argument assumes it without stating it directly, \hyp{LA} (Lean artifact) when it is introduced only by the formal encoding, and \hyp{PD} (potential drift) when it has no identifiable basis in the source.
Table~S3 in the supplement lists 35 conditions: 22 \hyp{HC}, 9 \hyp{IMP}, 3 \hyp{LA}, and one \hyp{PD}, which is the quantifier-order condition \hyp{H-NEW} mentioned in Section~\ref{sec:case3}.
Section~S6 of the supplement sets out the auditing protocol, including the tags used in the source files, the way in which the AI tools prepared the identifiers and tiers and how we checked them, as well as the rules according to which a condition can be included in a compiling theorem.


\begin{table}[t]
\centering
\small
\caption{Summary of the certification status of the five cases.}
\label{tab:cert-hierarchy}
\begin{tabular}{@{} >{\raggedright\arraybackslash}p{0.7cm} >{\raggedright\arraybackslash}p{4.6cm} >{\raggedright\arraybackslash}p{3.3cm} >{\raggedright\arraybackslash}p{4.0cm} @{}}
\toprule
Case & Kernel-checked core & Explicit theorem inputs & Outside the formal claim \\
\midrule
1 & Abstract argmax lemmas; likelihood failure certificates; moment-estimator consistency & iid, moments, integrability & Normal model and likelihood-argmax identification \\
2 & Dominated interchange; normal zero-mean score; varying-support example & Domination and measurability & Second score/Fisher-information identity \\
3 & Mean-value, uniform, and random-argument lemmas; the compiled non-implication; multiplicative Slutsky and the final assembly & Expansion, consistency-derived Hessian limit, score CLT & Model-specific MLE normality; a probabilistic uniform law of large numbers \\
4 & Differentiability chain, counterexample, deterministic curve lemma; the Slutsky wrapper & Remainder control and scalar weak limit & Probabilistic functional delta method and influence-function variance \\
5 & Deterministic oracle inequality and abstract union bound; sub-Gaussian maximal inequality & Two tail bounds & Random-design covariance concentration, measurable argmin, event measurability \\
\bottomrule
\end{tabular}
\end{table}

\subsection{Consistency of the MLE: the non-compactness obstruction}
\label{sec:case1}
Probably the first asymptotic result that a graduate student encounters is the consistency of the maximum likelihood estimator.
According to the law of large numbers, the sample expected log-likelihood, which is given by $M_n(\theta) = n^{-1}\sum_{i=1}^n \log f(X_i; \theta)$, converges to the corresponding population value $M(\theta) = \E_{\theta_0} \log f(X; \theta)$; since the model is identifiable, the function $M$ reaches its unique maximum at $\theta_0$ by the Wald information inequality; it follows that the maximizer of $M_n$ converges to the maximizer of $M$.
A separate theorem is required for the final step since the first two do not cause the maximizer of $M_n$ to converge to $\theta_0$.
The theorem is given by Wald's method \citep{wald1949consistency}.
The modern Wald-type statements such as \citet[Theorem~5.14]{vandervaart1998asymptotic} and \citet[Theorem~17]{ferguson1996course} call for a compact parameter set, an integrable upper envelope for the log-likelihood, upper semicontinuity with respect to $\theta$, and identifiability.
The conclusions that Wald reached in 1949 were based on different assumptions, and generalizations going beyond both of these approaches are classical \citep[see][]{huber1967behavior}.
Our work here is based on the modern framework, not on Wald's original statement.
As pointed out by \citet[Ch.~2, Example~3]{pollard2010asymptopia}, the consistency framework fails in two respects for the normal location-scale family on $\Theta = \R \times (0,\infty)$: the set $\Theta$ is not compact, and at $\mu = x$ the per-observation log-likelihood $\ell(x; \mu, \sigma) = -\log(\sqrt{2\pi}\,\sigma) - (x-\mu)^2/(2\sigma^2)$ tends to $+\infty$ as $\sigma \to 0$ and to $-\infty$ as $\sigma \to \infty$, which means that $\sup_\theta \ell(x; \theta) = +\infty$ and $\inf_\theta \ell(x; \theta) = -\infty$.
The unbounded-likelihood phenomenon is a classical one \citep{kiefer1956consistency}.
The result concerning consistency is still true, and Pollard's solution (that is, his Theorem~10 when it is applied to the normal family in Example~11) recovers it.

Since the Mathlib package in question does not include a Wald-type theorem, the development has to state and prove from scratch an argmax engine, \texttt{argmax\_consistency\_of\_wellSeparated\_uniform}, this engine deriving consistency from a well-separated maximum and uniform convergence.
The development does not formalize the compactness and envelope conditions from which uniform convergence would result, and the unrestricted normal model fails to meet these conditions.

\begin{figure}[t]
\leanfigure{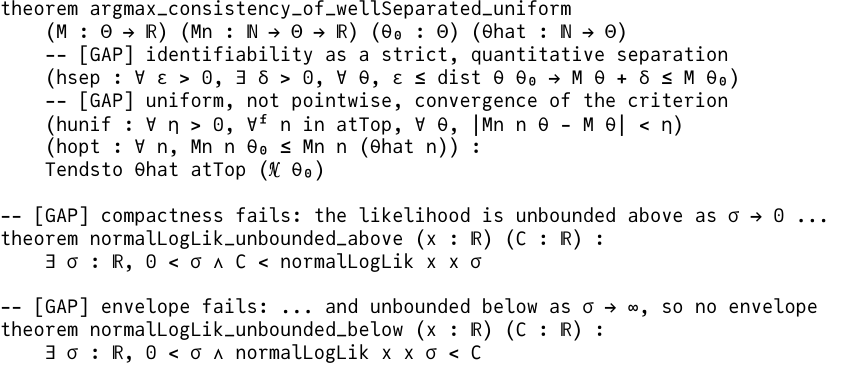}
\caption{Case~1: the consistency theorem and the two failure certificates.}
\label{fig:lean-01}
\end{figure}

The diagnosis was carried out by the \emph{statement} of the engine, since the machine imposed a choice at the two points where the informal argument is ambiguous.
Identifiability could not be expressed by the pointwise inequality $M(\theta) < M(\theta_0)$ when $\theta \neq \theta_0$ since the proof relies on a strict, quantitative separation: for every $\varepsilon > 0$ there exists a gap $\delta > 0$ such that $M(\theta) + \delta \leq M(\theta_0)$ provided that $d(\theta, \theta_0) \geq \varepsilon$.
This separation is equivalent to the sup-form well-separation condition stated in \citet[Theorem~5.7]{vandervaart1998asymptotic}, and the engine treats it as the hypothesis \texttt{hsep}.
The separated form can be deduced from the pointwise inequality if $\Theta$ is compact and $M$ is upper semicontinuous, but when $\Theta$ is non-compact the deduction may fail since $M(\theta)$ can approach $M(\theta_0)$ along a sequence that escapes every compact set.
Similarly, the statement ``by the law of large numbers'' could not be encoded as pointwise convergence of $M_n$, since the argmax step makes use of the uniform assertion $\sup_\theta \abs{M_n(\theta) - M(\theta)} \to 0$, which the engine takes to be the hypothesis \texttt{hunif}.
The transition from pointwise to uniform convergence is the step that makes use of compactness, the integrability conditions, and semicontinuity.
Table~S3 includes both condition \hyp{H1.4} and condition \hyp{H1.5} at the \hyp{HC} tier, the sources cited in their records stating these conditions.
Each item is associated with the step that uses it in the formalization.
The two failures were subsequently established as compiling theorems instead of merely being remarks, namely \texttt{normalLogLik\_unbounded\_above} and \texttt{normalLogLik\_unbounded\_below}.
What is being compiled has to be stated with precision.
The per-observation log-likelihood has no integrable envelope either above or below since it tends to infinity as $\sigma$ tends to $0$ at $\mu = x$ and tends to negative infinity as $\sigma$ tends to infinity, and therefore the compact-envelope framework cannot be applied as stated.
For a sample of size $n$ which is equal to or greater than 2 and which includes at least two different values, the sample log-likelihood still reaches its maximum at the closed form expressed by the sample mean and the uncorrected sample variance.
The engine and the two failure certificates are shown in Figure~\ref{fig:lean-01}.

The result regarding consistency is reached by following Pollard's Example~11.
The proof of \texttt{normal\_mle\_consistent} makes use of the strong law with respect to the first two empirical moments, thus rendering the weighted-comparison argument from Theorem~10 unnecessary.
The statement of the theorem applies to any sequence of pairwise independent and identically distributed variables which have a finite second moment, and it makes no assumption about normality nor about the closed form maximizing the likelihood.
The classical calculus derivation, which includes boundary analysis in $\sigma$, proves that the closed form is the maximum likelihood estimator for the normal model, although the development does not formalize it.
The identification thus defines the boundary of the formal statement.
The identification is located outside the formal claim in Table~\ref{tab:cert-hierarchy} and is recorded in the \Lean\ sources as the \texttt{[GAP]} \hyp{H1.7}, a condition which is not used as a hypothesis by any theorem.
The item \hyp{H1.7} is assigned to the \hyp{IMP} tier since the source mentioned in the record assumes the identification even though it does not state it directly.
The tier refers to that source, not to a missing condition in the classical result.
The case therefore divides the argmax step — a step which requires identifiability in its separated form and uniform convergence — from the closed-form approach, which does not need either of these but is based on identifying the closed form with the maximizer.

\subsection{Differentiation under the integral sign: what the interchange requires}
\label{sec:case2}
The two standard score identities are given in \citet[\S\S 2.5, 6.3]{lehmann1998theory}:
\[
\E_\theta[s(X;\theta)]=0,
\qquad
I(\theta)=\Var_\theta[s(X;\theta)]
          =-\E_\theta[\ell''(X;\theta)].
\]
Both results can be obtained by differentiating the normalization condition $\int f(x;\theta)\,dx=1$ with respect to $\theta$ and then exchanging the differentiation and integration.
The exchange has its own set of conditions and may fail when the support of $f(\cdot\,;\theta)$ varies as a function of $\theta$, as the counterexample at the end of this case demonstrates.
Explicit statements of the exchange require, for example, the domination of the derivative in a neighborhood of the parameter \citep[Theorem~2.4.3]{casella2002statistical}, or, as in the classical conditions of Cram\'er, a support which does not depend on the parameter.
The identities are important since a great deal of likelihood theory is based on them, including the information bound, the asymptotic variance of the MLE, and the expansion given in Section~\ref{sec:case3}.
At the exchange, the expression ``under suitable regularity conditions'' represents the conditions upon which the entire chain is based.

In this instance, in contrast to the other four cases, the Mathlib that is pinned already includes the required tool, the parametric-integral lemma \texttt{hasDerivAt\_integral\_of\_dominated\_loc\_of\_lip} \citep{mathlib2026parametric}.
The aim is to substitute the phrase with an already established and completely explicit theorem and to note down what the theorem requires that the phrase fails to include.

The Mathlib lemma was first placed inside the theorem \texttt{interchange\_under\_integral}.
The hypothesis list is the first result: this theorem necessitates the dominating function as an explicit object, together with a neighborhood of the parameter on which it dominates, and a proof that it is integrable.
The aforementioned theorem, \citet[Theorem~2.4.3]{casella2002statistical}, requires domination on a neighborhood \emph{and} the integrability of the dominating function, and the classical Cram\'er conditions play the same role.
We therefore checked that these conditions held in the case of a specific model, the normal location family, by constructing the dominating function \texttt{normalPDF\_dom\_bound} and proving that it was integrable.
We obtained the identity $\E_{\mu_0}[s(X; \mu_0)] = 0$ in two ways: by using the interchange (\texttt{normal\_score\_identity\_via\_interchange}) and also directly from the first moment of the Gaussian distribution (\texttt{normal\_score\_identity}).
The two results agree.
That agreement is what the exercise is about.
The fact that the identity was never open to question means that the calculation shows the interchange will achieve what the expression promises when its conditions are met.

We then set out the standard counterexample, this demonstrating the situation in which the conditions fail.
For the uniform family defined by $f(x; \theta) = \theta^{-1}\bm{1}_{[0,\theta]}(x)$, take any point $x_0$ that is in the interior.
The map which sends $\theta$ to $f(x_0; \theta)$ makes a jump from $0$ to $x_0^{-1}$ at the point where $\theta = x_0$, and so it is not even differentiable with respect to $\theta$ at that point.
The jump identifies the failure among the hypotheses listed for \texttt{interchange\_under\_integral}, since the condition of differentiability fails before the domination condition can be looked at.
The formal record includes four theorems.
The theorem \texttt{uniform\_not\_differentiable} demonstrates that differentiability fails.
The theorem \texttt{uniformPDF\_integral} shows that the parametric integral is always equal to $1$, which means its derivative is $0$; \texttt{uniform\_naive\_deriv\_integral} demonstrates that the integral of the interior derivative is $-\theta^{-1}$; and \texttt{uniform\_interchange\_fails} shows that these two values are different.

We believe that having a library of verified regularity conditions for the standard parametric families would enable a researcher to certify that the model in question satisfies the sufficient conditions required for a method of inference or an asymptotic approximation.
This shows for example why a proof assistant is extremely useful not only for theorist but also for those developing new methodologies and applications.
These tools allow to check and compare models and practical frameworks against the theory with high precision and rigor.

\subsection{Asymptotic normality of the MLE: the random-argument law of large numbers}
\label{sec:case3}
A first-order derivation of $\sqrt{n}(\hat\theta_n - \theta_0) \convd N\bigl(0, I(\theta_0)^{-1}\bigr)$ starts from the score equation $\psi_n(\hat\theta_n) = 0$ and expands it around $\theta_0$ by the mean value theorem:
\begin{equation}
\label{eq:mle-expansion}
0 \;=\; \frac{1}{\sqrt n}\sum_{i=1}^n \ell'(X_i; \theta_0)
\;+\; \left(\frac{1}{n}\sum_{i=1}^n \ell''(X_i; \tilde\theta_n)\right)\,
\sqrt n\,(\hat\theta_n - \theta_0),
\end{equation}
with $\tilde\theta_n$ an intermediate point between $\hat\theta_n$ and $\theta_0$.
Two limit steps finish the proof.
By the central limit theorem, the first term converges in distribution to $N(0, I(\theta_0))$.
By the law of large numbers, the bracket converges to $-I(\theta_0)$.
Slutsky's lemma then gives the conclusion, which \citet[Theorem~5.41]{vandervaart1998asymptotic} proves under classical conditions; see also \citet[Ch.~6]{lehmann1998theory}.
Van der Vaart's  proof of the classical-conditions statement handles the intermediate point explicitly, through a dominated argument involving the third derivative.
The step we examine is the second one.
The law of large numbers is a statement about a \emph{fixed} parameter value, while the bracket in \eqref{eq:mle-expansion} evaluates the empirical average at the \emph{random}, sample-dependent point $\tilde\theta_n$.
The step therefore needs a further result, convergence of the average at a random argument, which the law of large numbers does not deliver.
\citet[Chs.~16--18]{ferguson1996course} works through a uniform strong law, and \citet[Lemma~4.3]{newey1994large} state the needed random-argument lemma outright.

One scope choice was fixed in advance.
The CLT for the normalized score enters the final theorem as an explicit hypothesis, because the pinned snapshot contains no central limit theorem. 
As of October 2026, Mathlib proves the central limit theorem \citep{zhu2026clt}, but the development stays at the pin and keeps the hypothesis.
This case is also useful to understand what one can do when a needed result is not yet available from well established libraries. 
The formalized content is therefore the deterministic and strong-law part of the proof, which contains the step examined here.
The expansion step, \texttt{score\_equation\_expansion}, brings out two conditions at once.
First, the mean value theorem requires the score to be differentiable on the entire closed segment between $\theta_0$ and $\hat\theta_n$.
Since $\hat\theta_n$ may sit anywhere, this is a convexity requirement on the parameter set, although the expansion is centered at $\theta_0$.
Second, the starting point $\psi_n(\hat\theta_n) = 0$ is not free, since it presupposes that the estimator is eventually an interior stationary point.

The central finding concerns the LLN step.
When the formalization reached ``by the law of large numbers,'' it found no theorem to invoke.
The pointwise strong law has the shape
\[
\forall\, \theta,\;\; \forall^{\mathrm{a.s.}}\,\omega,\;\;
n^{-1}\textstyle\sum_i \ell''(X_i(\omega); \theta)
\;\longrightarrow\; \E\,\ell''(X; \theta),
\]
with $\theta$ bound \emph{outside} the almost-sure quantifier.
A theorem of this shape cannot be instantiated at a sample-dependent sequence $\tilde\theta_n(\omega)$, because that would require the parameter and the exceptional null set to vary together.
The file records only a deterministic version of it, \texttt{pointwise\_slln\_fixed}, a deliberately trivial, true lemma, which has no almost-sure quantifier and keeps only one order, $\theta$ fixed before the limit in $n$.
The lemma is plain function application and certifies nothing by itself, but it fixes, in checkable form, the order that the substitution of $\tilde\theta_n$ violates.
That the order matters is then a compiled statement, \texttt{pointwise\_not\_random\_arg}, which exhibits a family $f_n$ with $f_n(\theta) \to 0$ at every fixed $\theta$, and a sequence $a_n \to 0$, along which $f_n(a_n) = 1$ for every $n$.

The auditing protocol requires every condition used in a proof to be traced to a source.
We checked the sources for this case: van der Vaart \S 5.6, Ferguson Chs.~16--18, Lehmann--Casella Ch.~6, and Newey--McFadden, as recorded in the file header.
These pages give the repair a basis, in the uniform strong law of Ferguson and in Lemma~4.3 of Newey and McFadden, but none of them states the obstruction itself, namely that the pointwise law does not reach a random argument.
The audit therefore records the obstruction as \hyp{H-NEW}, at the \hyp{PD} tier.
The finding concerns a step, not a theorem, and no conclusion of these sources is false.
The citation ``the LLN'' names a theorem whose statement does not perform the step, and the machine accepts the substitution only from a theorem that does.

Mathlib has no uniform law of large numbers, so we proved a deterministic stand-in by hand, in dimension one: \texttt{ulln\_of\_pointwise\_equilipschitz}.
It is an equicontinuity lemma, which states that on a compact neighborhood $U \ni \theta_0$ an equi-Lipschitz family that converges pointwise converges uniformly.
Applied per $\omega$ to empirical criteria, it stands in for a uniform LLN.
A uniform law on $U$ needs an integrable envelope $\sup_{\theta \in U} \abs{\ell''(x; \theta)} \leq M(x)$, which the pointwise law does not require.
The development strengthens it to a single Lipschitz constant.
Uniform convergence on $U$, continuity of the limit at $\theta_0$, and consistency $\tilde\theta_n \to \theta_0$ then give the random-argument convergence, \texttt{random\_arg\_tendsto}, with the almost-sure form \texttt{random\_arg\_tendsto\_ae} carrying the probabilistic content used downstream.
The random-argument lemma is the analog of \citet[Lemma~4.3]{newey1994large}.
Consistency of $\tilde\theta_n$ enters as a hypothesis here.
Although the argmax theorems of Section~\ref{sec:case1} conclude a statement of the same form, the development does not compose the two.
The final theorem \texttt{asymptotic\_normality} assembles three inputs: the expansion, which holds for almost every $\omega$ eventually in $n$ (the estimator is eventually an interior stationary point); the almost-sure limit of the bracket; and the CLT hypothesis.
The CLT is carried abstractly as a limit law $\nu_S$.
Its intended instantiation $N(0, I(\theta_0))$ is centered because of the first score identity of Section~\ref{sec:case2}, which the development formalizes.
The positivity of the information is an explicit hypothesis, $0 < I$ (\texttt{hI}), which the closing step uses to invert the limit $-I$.

The conclusion is $\sqrt n\,(\hat\theta_n - \theta_0) \convd \nu_S$ rescaled by $I^{-1}$, under the probability-space and measurability hypotheses without which its integrals would be degenerate.
The snapshot proves Slutsky's theorem for random variables converging to a limit random variable on the same space, but not with the limit presented as a law, which is the only form in which $\nu_S$ reaches the statement.
The packaging that closes the gap is proved in \path|slutsky_packaging.lean| and related to the library's own predicate in \path|coverage_audit.lean|.
As of October 2026, Mathlib lets the limit be defined on a probability space of its own and states Slutsky's theorem for that predicate \citep{mathlib2026convdist}, so that a law can now serve as the limit.
The following fragment shows the obstruction and its repair side by side: the fixed-$\theta$ framework of the pointwise law, and the random-argument lemma with its two extra inputs (uniform control and continuity) in the type.
Its comment on the almost-sure quantifier refers to the law displayed above, not to \texttt{pointwise\_slln\_fixed}.
What the repair adds is readable off the second type, whose hypotheses \texttt{hunif} and \texttt{hcont} have no counterpart in the statement that ``by the law of large numbers'' cites.

\subsection{The influence function: G\^ateaux versus Hadamard differentiability}\label{sec:case4}
A statistical functional $T$ is estimated by the plug-in $T(\hat F_n)$, and its asymptotic behavior is studied via the von Mises expansion
\begin{equation}
\label{eq:vonmises-expansion}
T(\hat F_n) - T(F) \;=\; \int \mathrm{IF}(x; T, F)\, d(\hat F_n - F)(x) \;+\; R_n,
\end{equation}
where the influence function is the derivative of $T$ at $F$ with respect to the contamination proportion: $\mathrm{IF}(x; T, F) = \frac{d}{dt} T\bigl((1-t)F + t\delta_x\bigr)\big|_{t=0^+}$ \citep{hampel1986robust,huber2009robust}.
If $R_n = o_P(n^{-1/2})$, then the linear term is subject to the central limit theorem and we have $\sqrt n\,\bigl(T(\hat F_n) - T(F)\bigr) \convd N\bigl(0, \int \mathrm{IF}^2\, dF\bigr)$.
Everything therefore hinges on the negligibility of $R_n$, and that in turn depends on what kind of differentiability property $T$ possesses.
The influence function is a directional (G\^ateaux-type) object.
The functional delta method (controlling $R_n$ along the random paths $\hat F_n$) requires Hadamard (compact) differentiability \citep[Ch.~20]{vandervaart1998asymptotic}.
Hadamard differentiability is strictly stronger than the directional notion \citep[see][]{reeds1976vonmises,fernholz1983vonmises}, in that it controls the difference quotients along directions that vary with $n$ and converge, not just along fixed directions.
The statement that $T$ is differentiable with influence function $\mathrm{IF}$ has two readings and computing $\mathrm{IF}$ will establish at best the directional one, which is not sufficient to control $R_n$.

This one was the most difficult of the five, as the other cases were all based on notions which were already in place.
Here we had to define the notions themselves, and in a proof assistant a definition is a design act.
The definitional layer on its own produced three findings.
Firstly the G\^ateaux derivative was defined with its candidate derivative as a bare function $L : E \to \R$.
G\^ateaux differentiability does not give linearity or continuity of $h \mapsto L(h)$, and with a bare function this lack is visible.
The integral form $\int \mathrm{IF}\, d(\cdot)$ of the derivative has both properties as an assumption rather than as a consequence.
Secondly the von Mises derivative was defined as a one-sided affine notion along mixtures $F + t(G - F)$, $t \downarrow 0$, with $G$ ranging over a convex set, in the spirit of \citet{cerreiavioglio2024affine}.
The reason for this is that the distribution functions form a convex set rather than a vector space and $F + t(G-F)$ is in general only a distribution function for $t \in [0, 1]$.
Thirdly the ambient norm had to be declared prior to any statement being written.
Sup-norm and total variation lead to different differentiability classes so the norm is part of the statement.
The second of those findings has an implication.

Hadamard differentiability, the implication chain and the curve lemma are all stated for a functional defined on the whole normed space, since every function in \Lean\ is defined on its entire domain type.
A statistical functional on the other hand is defined only on the convex set of distribution functions, and only the von Mises notion is restricted to that set.
The implicit extension of $T$ to the whole space is recorded as a Lean-artifact condition, together with the observation that the admissible sequences $t_n \to 0$ of the Hadamard definition may change sign.
That change of sign is innocuous on a vector space and inadmissible on a convex set of distribution functions.
The implication chain Fr\'echet $\Rightarrow$ Hadamard $\Rightarrow$ G\^ateaux $\Rightarrow$ von Mises was then established.
The heart of the case is the strictness of the middle link, which we established via a compiled counterexample.
The ``moving parabola'' function on $\R^2$ is $1$ on the punctured parabola $\{b = a^2,\, a \neq 0\}$ and $0$ elsewhere.
It is G\^ateaux differentiable at the origin with derivative $0$, as every fixed line intersects the punctured parabola at most once.
It is not Hadamard differentiable there (along $h_n = (1, t_n)$ with $t_n = 1/(n+1)$ the perturbation $t_n h_n$ hits the parabola exactly and the difference quotient $1/t_n$ diverges).
Both halves compile as \texttt{parabola\_gateaux} and \texttt{parabola\_not\_hadamard}.
The machine-checked conclusion is that G\^ateaux differentiability does not license the delta method.

\begin{figure}[H]
\leanfigure{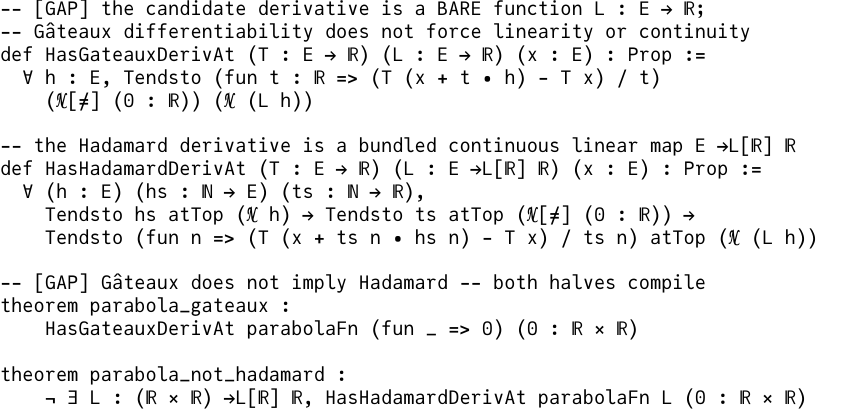}
\caption{Case~4: G\^ateaux and Hadamard separated at the level of types.}
\label{fig:lean-03}
\end{figure}

We then split the delta method up and what each half contains is part of the finding.
The analytic half is fully proved via \texttt{hadamard\_delta\_method\_curve}: if $T$ is Hadamard differentiable at $F$ and a deterministic curve has norm-convergent rescaled increments $t_n^{-1}(F_n - F) \to G$, then $t_n^{-1}(T(F_n) - T(F)) \to L(G)$.
With $t_n = 1/\sqrt n$ this is the statement that the remainder in \eqref{eq:vonmises-expansion} vanishes after scaling along such a curve.
In the probabilistic half the certification boundary requires care.
Donsker-type weak convergence of $\sqrt n(\hat F_n - F)$ does not provide almost-sure path control, so the curve lemma cannot be applied to the empirical process $\omega$ by $\omega$.
The classical bridge is the almost-sure representation and the extended continuous-mapping argument of \citet[Ch.~20]{vandervaart1998asymptotic}, which we did not formalize.
The distributional object \texttt{functional\_delta\_method} therefore takes two explicit inputs, under the same scope discipline as the CLT in Section~\ref{sec:case3}.
The first is a scalar central-limit hypothesis for the linearized statistic $t_n^{-1} L(\hat F_n - F)$, the one-dimensional consequence of a Donsker input rather than the process-level statement.
The second is the almost-sure remainder hypothesis which serves as a stand-in for the representation argument.
The finite variance which makes the limit law well defined is recorded as the implied square integrability of the influence function.
The final Slutsky step  uses the lemma of Section~\ref{sec:case3}.
Figure~\ref{fig:lean-03} separates the two notions at the level of types, with the G\^ateaux candidate derivative being a bare function and the Hadamard derivative a bundled continuous linear map.

\subsection{The LASSO oracle inequality: coupling deterministic and probabilistic arguments}\label{sec:case5}
Consider the sparse linear model $Y = X\beta_0 + \varepsilon$, where $X \in \R^{n \times p}$ with $p \gg n$ and $\beta_0$ is supported on $S_0$ with $\abs{S_0} = s$.
The LASSO estimator $\hat\beta \in \argmin_\beta \tfrac{1}{2n}\norm{Y - X\beta}_2^2 + \lambda \norm{\beta}_1$ satisfies an oracle inequality.
When $\lambda \asymp \sigma \sqrt{\log p / n}$ and the restricted-eigenvalue condition holds on the cone $\mathcal{C}(S_0, 3)$ with constant $\kappa$, the estimation error satisfies $\norm{\hat\beta - \beta_0}_2 \leq 3\lambda\sqrt{s}/\kappa$ with high probability \citep[Theorem~7.13]{wainwright2019high}.
The $\ell_2$/restricted-eigenvalue analysis goes back to \citet{bickel2009simultaneous}, the companion prediction/$\ell_1$ oracle inequality under the compatibility condition can be found in \citet[Ch.~6, Theorem~6.1]{buhlmann2011statistics}, and a survey of the general sparsity theory is given by \citet{vandegeer2016estimation}.
The chain of the formal argument can be summarized as follows:
the basic inequality is derived from the minimality of $\hat\beta$ \citep[eq.~(6.3)]{buhlmann2011statistics}; on the event $\norm{X^\top \varepsilon / n}_\infty \leq \lambda/2$ the noise term is bounded by means of H\"older's inequality, which forces the error into the cone \citep[Lemma~6.3]{buhlmann2011statistics}; the restricted-eigenvalue condition then turns the prediction error into an estimation error.
In contrast to Sections~\ref{sec:case1}--\ref{sec:case4}, this case deals with modern finite-sample theory.

The argument was split into a deterministic part and a probabilistic part, and the two halves fared very differently.
The deterministic chain is fully proved, whereas the probabilistic part states explicit events, takes the two tail bounds as assumptions, and proves only their coupling by a union bound.

The deterministic part is considered first.
Firstly, the notation $\hat\beta = \argmin(\cdot)$ presupposes a minimizer, which does exist but need not be unique when $p > n$, so that a probability statement about $\hat\beta$ requires a measurable choice to be made among the minimizers.
Since the deterministic chain needs nothing more than the minimality inequality, this inequality was taken as a hypothesis.
Secondly, the unknown support had to enter the statement.
Both the cone and the restricted-eigenvalue constant depend on $S_0$, which makes ``RE holds'' a family of conditions quantified over supports; the formal theorem therefore carries $S_0$ as an explicit finite set, together with the sparsity witness.
Thirdly, the constants had to be tracked as objects.
The opening $3$ of the cone is determined by the choice of $\lambda/2$, and $\lambda > 0$ has to be assumed.
Fourthly, and this is the most instructive of the four findings, the normalization of the restricted eigenvalue had to be declared.
Under the condition $\tfrac1n \norm{X\Delta}_2^2 \geq \kappa^2 \norm{\Delta}_2^2$ on the cone, the bound which is actually proved reads
\begin{equation}
\label{eq:lasso-honest}
\norm{\hat\beta - \beta_0}_2 \;\leq\; \frac{3\lambda\sqrt{s}}{\kappa^2}.
\end{equation}

In the deterministic chain the deviation and the restricted-eigenvalue conditions are predicates on fixed data; from the probabilistic point of view, on the other hand, they become events on a probability space.
The passage from the one to the other is made explicit by the coupling theorem, in which the data and the estimator are taken as functions on a single probability space and each condition as the set of outcomes at which it holds.
The intersection of the deviation event and the restricted-eigenvalue event, that is, the set of data realizations on which both conditions hold, will be called the \emph{good event}.
On this event the bound \eqref{eq:lasso-honest} is delivered pointwise by the deterministic chain.
The probability of the good event is bounded from below by a union bound on the two failure probabilities and not by a product of marginal probabilities, since the two events are not necessarily independent.
This coupling is proved as \texttt{lasso\_oracle\_union\_bound}, a theorem whose hypotheses are the two tail bounds themselves: $\Prob(\mathrm{DEV\ fails}) \leq c_1 p^{-1}$, which sub-Gaussian noise yields under the calibration $\lambda \asymp \sigma\sqrt{\log p/n}$, and $\Prob(\mathrm{RE\ fails}) \leq c_2 e^{-c_3 n}$, which likewise follows from sub-Gaussianity.
The conclusion of the theorem is that the oracle bound holds with probability at least $1 - c_1 p^{-1} - c_2 e^{-c_3 n}$.
As no measurability is assumed, the bound concerns the outer probability of that event, which coincides with its probability in the case where $\hat\beta$ is measurable.

The deviation bound is a maximal inequality over the $p$ coordinates.
The maximal inequality for a finite family of sub-Gaussian variables, together with its sum form $2p\,e^{-\varepsilon^2/(2nc)}$, is proved in \path|coverage_audit.lean| from the Chernoff and Hoeffding bounds of the snapshot.
What is still missing is the design side, namely covariance concentration and the verification of the restricted-eigenvalue condition for random designs.
The restricted-eigenvalue bound enters as a hypothesis, in accordance with the policy that was followed for the CLT in Section~\ref{sec:case3} and for the Donsker input in Section~\ref{sec:case4}.
According to that policy, whatever the library cannot yet supply is assumed explicitly rather than proved.
The deviation bound enters in the same form, so that the two inputs of the coupling remain symmetric.
As of October 2026, part of that infrastructure is available in libraries outside Mathlib (Section~\ref{sec:background}).
The matrix Bernstein inequality, as well as a two-sided bound on the singular values of random matrices whose rows are independent, centered, isotropic and sub-Gaussian, is proved in the repository of \citet{zhang2026ai4slt}.
StatLean proves the deterministic chain up to the prediction and $\ell_1$ error bounds, while matrix Bernstein is stated there as an axiom \citep{statlean2026}.
Neither of them proves the restricted-eigenvalue condition for random designs with $p > n$, and neither is built on the pinned Mathlib, which is why the development keeps both hypotheses.

An earlier version of this case contained a mistake which shows the way in which the development strategy can go wrong.
The two tail bounds were first stated as separate admitted theorems, each of them claiming ``there exist constants $c_1, c_2, c_3$ such that the bound holds,'' at fixed $n$ and $p$.
The file compiled, and nothing but the two admitted steps was flagged by the checker.
We then noticed that these statements are nearly empty: with $n$ and $p$ fixed, any probability whatsoever is bounded by $c_1 p^{-1}$ as long as $c_1$ can be chosen freely.
With $c_1 = p$, for instance, the bound reads $\Prob \leq 1$.
Proofs of these statements would have compiled as well, since the statements are true; they are simply not the statements which the LASSO theory needs.
The probabilistic part was therefore redesigned.
The constants $c_1, c_2, c_3$ are now fixed once and for all in the hypotheses of the coupling theorem, and the same constants reappear in the conclusion, so that they cannot be chosen in such a way as to make the hypotheses trivial.
The lesson is the recurring theme of this paper seen from the opposite side: the machine checks that a statement has been proved, but whether the statement says anything at all can only be checked by a human.

\section{Lessons for Research and Advanced Teaching}
\label{sec:synthesis}
The five cases were chosen to exhibit mechanisms rather than to prove new theorems, and this section draws the lessons from them.
The first question is which kind of conditions the formalization brought out and at which steps they enter the arguments; the others concern where the work of verification moves once a kernel checks the proofs, how a development of this kind is produced, and what follows for our understanding of the theoretical foundations,  doctoral training and for a library of statistical theory.

\paragraph{\bf Patterns.}
The three \hyp{LA} conditions of Table~S3 are set aside here.
Of the remaining 32, the 22 at the \hyp{HC} tier are not missing hypotheses which the formalization discovered: each of them ties a condition that the source states to the step of the argument which uses it, a link that a compressed citation may miss.
Of the ten conditions that the sources leave unstated, four (\hyp{H1.7}, \hyp{H4.6}, \hyp{H5.1}, and \hyp{H5.8}) have no compiled counterpart and are listed by Table~S4 as prose-only limitations or interpretations, while the other six are carried by compiled declarations, five as theorem hypotheses and \hyp{H-NEW} as the non-implication of Section~\ref{sec:case3}.
The 32 conditions sort into four mechanisms, which Table~\ref{tab:taxonomy} names and locates in the cases: the geometry of the parameter set, the analytic control of envelopes and differentiability classes, the quantifiers over randomness, and the bookkeeping of constants and normalizations.

\begin{table}[t]
\centering
\caption{Taxonomy of the gaps found in the five formalizations.}
\label{tab:taxonomy}
\begin{tabular}{@{}p{2.5cm}p{6.1cm}p{3.6cm}@{}}
\toprule
Mechanism & What is at stake & Cases \\
\midrule
Topological & compactness, convexity, interiority, convex domain geometry & \S\ref{sec:case1}, \S\ref{sec:case3}, \S\ref{sec:case4} \\
\addlinespace
Analytic & envelopes, domination, support, differentiability class, norm choice & \S\ref{sec:case1}, \S\ref{sec:case2}, \S\ref{sec:case3}, \S\ref{sec:case4} \\
\addlinespace
Probabilistic & quantifiers over randomness, couplings, joint events & \S\ref{sec:case3}, \S\ref{sec:case4}, \S\ref{sec:case5} \\
\addlinespace
Algebraic & constant tracking, normalization conventions & \S\ref{sec:case5}\\
\bottomrule
\end{tabular}
\end{table}

Three patterns cross the boundaries between cases.
The \emph{domination/envelope} mechanism appears three times: as the integrable envelope which the unrestricted normal family lacks (Section~\ref{sec:case1}), as the dominating function which the exchange of derivative and integral requires as an explicit, integrable object (Section~\ref{sec:case2}), and as the envelope on a neighborhood of $\theta_0$ which a uniform law needs and the pointwise law does not (Section~\ref{sec:case3}); in all three the condition is used and applied at a step that the informal argument attributes to a limit theorem or to regularity conditions.
The \emph{coupling} of a deterministic with a probabilistic argument appears in the remainder of the delta method (Section~\ref{sec:case4}) and in the joint good event of the LASSO bound (Section~\ref{sec:case5}).
The third pattern cuts across several gaps at once: pointwise versus uniform convergence (Section~\ref{sec:case1}), fixed versus random argument (Section~\ref{sec:case3}) and the quantifier over the unknown support (Section~\ref{sec:case5}) are all questions of \emph{quantifier order}, which a sentence in natural language can leave unmarked and which a formal statement fixes by the order of its binders (Section~\ref{sec:primer}).
Quantifier order was the issue we met most often within this development, and it is also the one that a working theorist is least trained to watch.
The experience with the elaboration of these classical results using the proof checker is generally that we understood more and we were capable of producing more robust formal arguments.
Moreover, some of these proof strategies that are now less affected by the risk of making subtle mistakes that can compromise a long chain of formal results, can be reused in structured way. 

\paragraph{\bf Evidence from independent formalizations.}
The five cases were chosen to exhibit these mechanisms and their tiers were assigned within this project, so that the counts above describe this development and say nothing about how widespread the mechanisms are.
Findings of the same kind are, however, reported by three projects which formalize statistical theory at a larger scale, with other aims and independently of this paper; the independence is partial in one respect only, since our tier scheme is the one of \citet{wei2026hypothesis} (Section~\ref{sec:cases}), so that the two inventories share a classification though not a case or a source.
\citet{wei2026hypothesis} formalize the asymptotic theory of parametric and semiparametric models from \citet{vandervaart1998asymptotic}, and every hypothesis of their five main theorems is tied by their audit to a quoted passage and a page of that book.
No unsupported hypothesis is found on any main theorem \citep[Sections~3.4 and~A.3]{wei2026hypothesis}, but the local asymptotic minimax bound shows a condition which the source states away from the theorem it serves \citep[Section~3.5]{wei2026hypothesis}: in its opening sentence, outside the statement of the theorem, the proof of Theorem~8.11 of van der Vaart restricts itself to lower semicontinuous losses and uniformly tight estimators, and both restrictions are carried by the formal signature as hypotheses, so that the formal theorem is the one which the published proof establishes.
In the same development, four unsupported hypotheses were added by agents under proof pressure to candidate statements of that theorem, and the file still compiled with no admitted step and the standard axioms only; the audit, and not the kernel, removed them \citep[Appendix~B.1]{wei2026hypothesis}.
\citet{zhang2026ai4slt}, who formalize empirical-process tools for statistical learning theory, report that their formal version of the master error bound of \citet[Theorem~13.5]{wainwright2019high} carries as explicit hypotheses the integrability of the supremum of the empirical process at two radii and the non-emptiness of an empirical sphere \citep[Appendix~C]{zhang2026ai4slt}, and that three misformalized target statements survived repeated checks by their AI agents and were caught by a human only through the construction of counterexamples \citep[Section~5.3]{zhang2026ai4slt}.
\citet{tan2026causalforge}, whose library and agent pipeline produce new results in causal inference, describe two failures from their own runs, an agent which turned difficult lemmas into axioms and a model which altered a hypothesis so that the code proved a statement other than the one claimed in the paper; the kernel accepted both \citep[Section~1]{tan2026causalforge}.

\paragraph{\bf Where verification moves.}
A formalization separates what a theorem \emph{asserts} from what its argument \emph{authorizes} (Section~\ref{sec:checker}), and in doing so it moves trust from the social, where everyone knows the step, to the mechanical.
The conclusions of the five arguments are true and the citations point to real theorems; what the formalization adds is an account of which condition each step uses.
Compression makes a proof legible, since a reader who knows the standard theory package does not need to have it rehearsed, and a textbook which expanded every regularity condition at every use would teach less rather than more.
The cost of the idiom is not error but non-recoverability.
In \citet[Ch.~6]{lehmann1998theory}, as in the derivation that accompanies \citet[Theorem~5.41]{vandervaart1998asymptotic}, the average at the intermediate point of \eqref{eq:mle-expansion} is said to converge by the law of large numbers; the condition which makes the step valid appears elsewhere, among the hypotheses of the theorem, and the citation at the step does not say which hypothesis the step uses.
From the compressed citation alone that link cannot be reconstructed, by a reader or by a referee.

When proofs are produced with AI assistance and certified by a kernel, the proof steps no longer need a human reader, and human verification moves to the statement and its hypotheses, which is exactly where the idiom compresses.
The kernel does not check that a formal statement renders the intended claim, and the cases show that this check carries mathematical content of its own: the tail bounds first stated in Case~5 were true, compiled and nearly empty, and only a reading of the statements exposed them (Section~\ref{sec:case5}).
The three independent projects reach the same point from three directions, with Wei et al.\ auditing every hypothesis against the source, Tan and Syrgkanis every formal statement against the informal claim, and Zhang et al.\ constructing counterexamples by hand.
Outside statistics, the Navier--Stokes release of Section~\ref{sec:intro} met the same objection: by the account of \citet[Section~3]{bastounis2026lost}, its \Lean\ theorem establishes less than the paper claims.
The conclusion we draw is not that statistical prose should be written like a \Lean\ file, but that the compressed idiom is a bibliographic device rather than a proof step, and that a discipline holding an instrument that can expand it on demand may keep the compression and still know what it conceals.
How widespread the compression is in the statistical literature is an empirical question, which five cases cannot answer and which audits on a corpus of textbooks and papers, of the kind that Wei et al.\ carried out on a single book, could; we leave that measurement as an agenda.
The cases also suggest a division of labor among three parties: the researcher states the claim, chooses the strategy and reads the hypotheses, a generative model proposes formal statements and candidate proofs, and the kernel checks each candidate, its rejections exposing the assumptions and definitions that the argument leaves implicit.
Neither machine can take over the first role, since a claim that compiles may still differ from the claim which the researcher meant.

\paragraph{\bf Producing the development.}
The \Lean\ sources were produced with three AI tools, Gemini~3.1~Pro, Claude Opus~5 and Leanstral~1.5, under our direction (see the declaration at the end of the paper), following the loop of Section~\ref{sec:background}: the tools generated and repaired \Lean\ code against the error messages of the compiler, while we chose the claims and read the statements that the tools produced.

A reliable time for this development cannot be given, since it did not start from scratch and concerned results which we knew in detail; on new problems that we are now developing with the same kind of tools, our estimate, which we have not measured, is a reduction of working time by a factor close to ten.
Two failure modes deserve a record, one caught by reading and one by the machine.
The first is the pair of tail bounds of Case~5, described above.
The second is the pair of closing Slutsky steps of Cases~3 and~4, admitted in the first version because the library held the needed results only in a form that the statements could not use; they were reported by the axiom profile and are now proved by \path|slutsky_packaging.lean|.

\paragraph{\bf Pedagogy.}
Graduate training in mathematical statistics teaches the standard steps of these proofs, expand, invoke, conclude, and the five cases show where each of those steps draws on a condition which the citation leaves implicit.
We believe that the combination of proof assistant and generative AI is an incredible opportunity for graduate student to learn more efficiently avoiding ambiguities from the beginning.
A student can see almost immediately how different proof strategies (the tactics) for the same statement can be developed based on the large catalog of literature that has been accumulated over decades of research.
Mistakes are quickly signaled by the kernel, and a researcher can better understand the details of the formal environment from her/his own mistakes with the help of AI tools aiding the formalization process.
Even if knew already a lot of details about the five cases presented in this paper, we learned a lot. 
Each case already contains an exercise for an advanced course, the two failure certificates of Section~\ref{sec:case1} or the moving parabola of Section~\ref{sec:case4}, for instance.
The available studies of \Lean\ in teaching concern first-year undergraduates and rest on small, self-selected samples; they report a gain in the precision of mathematical language and in the students' view of rigor, and a cost in time and syntax which kept most students from engaging with the system \citep{thoma2022lean,iannone2024perceptions,bottoni2025teaching}.
We know of no study at the doctoral level, where our proposal sits: a module in which each student formalizes the \emph{statement}, rather than the proof, of one theorem from their own reading list.
Our own experience in producing the development, anecdotal as it is, was that most of the effort went into fixing the framework in which an argument takes place rather than into the proof technique, because a formal environment does not let the framework stay unspecified; whether that shift benefits doctoral students is a question for a study of the kind cited above.

\paragraph{\bf Infrastructure.}
We believe that an important long-term goal is a community library for statistics, comparable in scope and role to Mathlib; the closest existing effort is Statlib, a community project which settles definitions and statements before proofs (Section~\ref{sec:background}; \citealp{statlib2026}).
The five formalizations of this paper are not a library, since they were designed to analyze five arguments, but three design choices for one can be read off them.
A result which the library lacks should enter as a named hypothesis of the theorem that uses it, as the central limit theorem does in Case~3, the Donsker input in Case~4 and the tail bounds in Case~5, and never as an axiom, which at the point of use cannot be told apart from a proved theorem (Section~\ref{sec:primer}).
\citet[Section~6.3]{tan2026causalforge} carry a published lower bound as an explicit hypothesis of their main result, so that a reader sees which part is proved and which is imported, whereas StatLean declares such results as axioms \citep{statlean2026}.
Convergence in distribution should be encoded once.
The cases needed a limit given as a law, which the pinned predicate did not accept, and the packaging of Case~3 fills that gap; as of October 2026, however, Mathlib has changed the predicate itself \citep{mathlib2026convdist}, and \citet[Appendix~C.3]{wei2026hypothesis} define their own weak-convergence predicate on measures, so that three encodings of one notion were produced within a few months, which shows the fragmentation that a shared vocabulary has to prevent.
A gap inventory of the form of Table~S3, which records each condition with its tier and the source cited for it, could be kept by a library for every theorem it takes from the literature, as Wei et al.\ do with their rows of source quotations.
Measured against the five cases, the priorities are a uniform law of large numbers for classes with an integrable envelope together with the Wald-type consistency theorem that rests on it (Cases~1 and~3), the functional delta method for statistical functionals on the convex set of distribution functions, with Hadamard differentiability (Case~4), and covariance concentration with the restricted-eigenvalue bound for random designs (Case~5).
Mathlib contains none of the three at the date above, although the central limit theorem and Slutsky's theorem for a limit given as a law, which the cases also needed, are now in it \citep{zhu2026clt,mathlib2026convdist}; parts of the first two exist outside Mathlib, since Causalean formalizes Glivenko--Cantelli tools, M- and Z-estimation and the delta method with Hadamard differentiability \citep[Sections~2 and~4.2]{tan2026causalforge}, while the third is proved by none of the libraries reviewed in Section~\ref{sec:background} (Section~\ref{sec:case5}).

Such a library would also make possible a form of reproducibility that theoretical statistics lacks at present, since a proof cannot be rerun and can only be inspected, typically by a referee.
An author could submit, alongside the paper, the \Lean\ source files whose compilation certifies the formal statements and proofs, so that their correctness would be checked by the kernel and would no longer have to be reconstructed from the informal argument.
Peer review would not be eliminated by such a submission; it would move to the two questions that the formal system cannot settle, whether the formal statement faithfully represents the claim made in the paper, a matter of human judgment which, as the cases show, can carry mathematical content of its own, and whether the result matters, in what it means, how it relates to existing results and what it contributes to the field.
This form of reproducibility is not imminent, as it requires the infrastructure described above and changes in the practices of authors, editors and referees, but it is a direction that the five cases point to, and a reason for statisticians, and not only formalizers, to treat formalization as part of the infrastructure of theoretical research.

\section{Conclusion}
\label{sec:conclusion}

The five cases were formalized in order to find out what a machine reader recovers from a compressed argument, and the answer is less dramatic, and more useful, than the talk of hidden assumptions suggests.
The conclusions of the cited sources are true; what the formalization adds is the link between each condition and the step which uses it, compiled certificates of the failures that the informal idiom passes over, and, in one case, a condition with no basis in the sources at all, the quantifier order at the random intermediate point of Section~\ref{sec:case3}.
The tail bounds of Section~\ref{sec:case5}, which compiled and said almost nothing, supply the other half of the lesson: the kernel certifies the proof of a statement, and only a reader can tell whether the statement says anything.
Three larger formalization projects met the same two findings on their own material (Section~\ref{sec:synthesis}), which carries the lesson beyond our cases.
 
The change that these tools ask of us is not a shift from pencil and paper, or chalk and blackboard, to a computer session; it is a change of mentality.
The theorist's work moves to the point at which the kernel stops, stating the claim, choosing its conditions and judging whether a formal statement says what the paper claims, and we found that work more instructive than the derivations it replaces, not less: most of our effort went into the frameworks in which the five arguments take place, and that is where the understanding was gained.
The same holds in the classroom, where a student who states a theorem discovers what the statement needs before any proof is attempted.
Nor is the gain confined to theory, since a shared formal library would let a methodologist or an applied statistician check, rather than assume, that a model falls within the scope of the theory whose guarantees are being invoked.
 
Much remains open.
Whether formal tools, with generative models now supplying the proofs, will make theory more reliable and its teaching clearer, and will let theorists try conditions and strategies on a scale that derivation by hand never allowed, is a conjecture; at the doctoral level, where our teaching proposal sits, no study that tests it is known to us.
The infrastructure is not there yet either: a library built on the priorities of Section~\ref{sec:synthesis} would let authors submit the \Lean\ sources of their theorems with their papers and leave referees to judge the statements.
 However the field answers these questions, the job has already changed, and doctoral training will have to teach the new one, stating claims and reading hypotheses, as carefully as it now teaches derivations.

\section*{Supplementary Material}
\addcontentsline{toc}{section}{Supplementary Material}

\noindent\textbf{Supplement to ``Statistical Theory in the Age of Machine-Assisted Mathematics''} \citep{coretto2026supplement}.

The supplement is the archival companion to this paper.
It contains the complete, version-locked \Lean~4 / Mathlib development.
It specifies the pinned environment and build procedure, explains what a successful build does and does not certify, reports the coverage audit against the pinned Mathlib commit, and provides a guide to the individual modules.
The supplement also contains the auditing protocol, the complete inventory of gaps, and the complete source code.
The source files are included in the supplement's \path|code/| directory and reproduced in its listings appendix.
A public repository holds a copy of the archive at
\begin{center}
\href{https://github.com/pietro-coretto/lean-matstat-case-studies}%
{\tt https://github.com/pietro-coretto/lean-matstat-case-studies}
\end{center}


\section*{Data Availability Statement}
This study generated and analyzed no datasets.
The \Lean~4 / Mathlib development that supports its findings is available in the supplementary material of this article, and the public repository given above holds a copy, under the Apache License~2.0.

\section*{Declaration on the use of artificial intelligence}
Artificial-intelligence (AI) systems were used in this work, in two roles that are disclosed separately.
All creative and intellectual content belongs to the author: the conception of the study, the choice and design of the five cases, and every mathematical and editorial judgment.

Development of the \Lean\ sources.
The sources were produced with ``Gemini 3.1 Pro'' (Google), ``Claude Opus 5'' (Anthropic), and ``Leanstral 1.5'' (Mistral AI), through an extended sequence of structured iterations directed and reviewed by the author.
No single prompt could reproduce the result.
The same tools drafted the initial \texttt{[GAP]} records and their tiers, which the author then checked against the cited sources (Section~\ref{sec:cases}).
The ten source files total 2{,}840 lines.
The author cannot give a reliable time for this development, which did not start from scratch and concerned results he knew in detail.
Section~\ref{sec:synthesis} reports his unmeasured estimate for new problems and the failure modes he found.

Preparation of the manuscript.
The author combined ``Claude Opus 5'' (Anthropic), ``Kimi K3'' (Moonshot AI), and ``GPT-5.6-Luna'' (OpenAI) for drafting, code refactoring, language editing, and consistency checking of the text.

After using these tools, the author reviewed and edited all content and takes full responsibility for the content of the publication.
No artificial-intelligence system meets the criteria for authorship, and none is credited as an author of this work.

\bibliographystyle{chicago}
\bibliography{REFS}

\includepdf[pages=-]{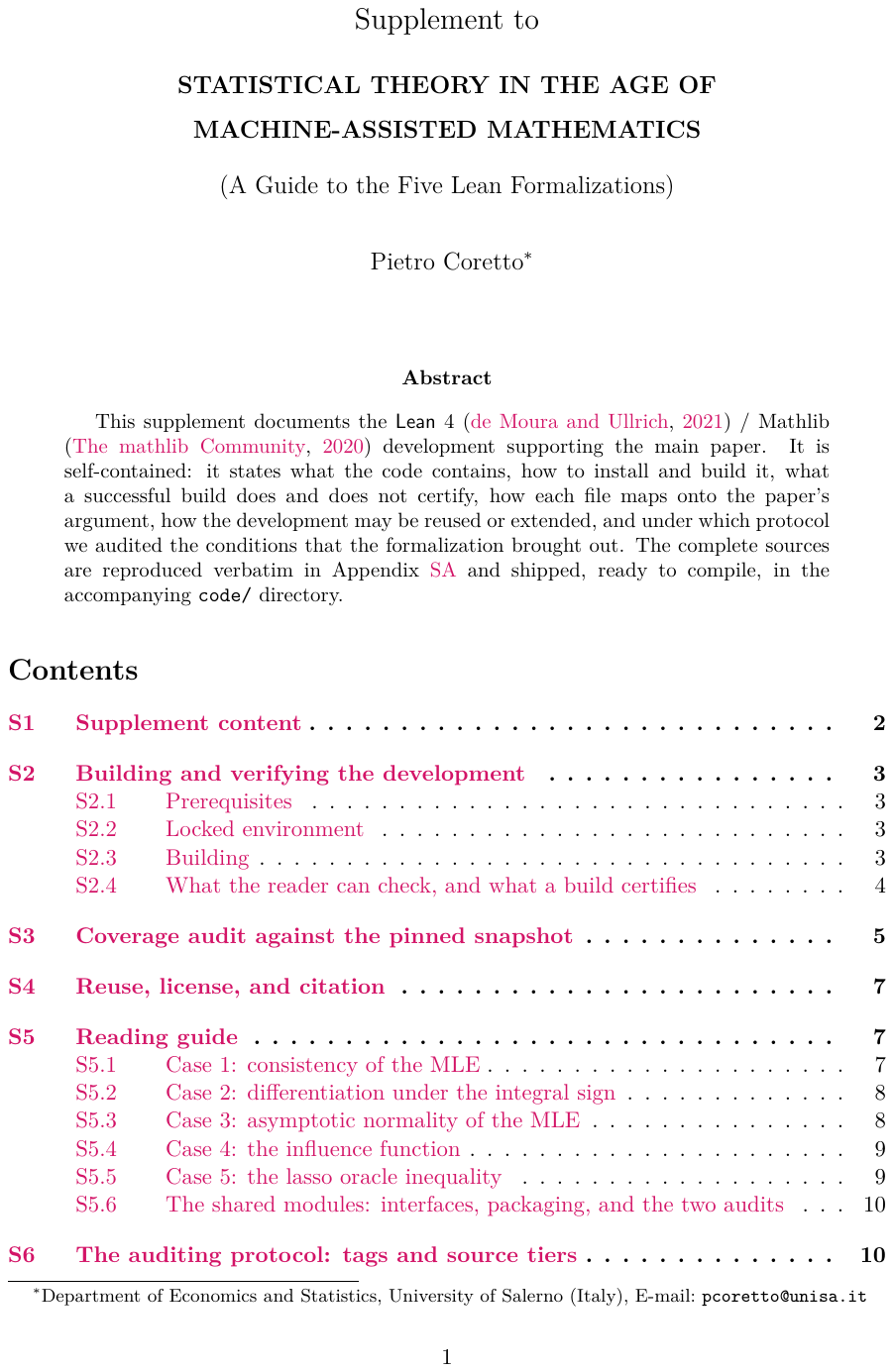}
\end{document}